\documentstyle[12pt]{article}

\newbox\SlashedBox  
\def\fs#1{\setbox\SlashedBox=\hbox{#1} 
\hbox to 0pt{\hbox to 1\wd\SlashedBox{\hfil/\hfil}\hss}{#1}} 
\def\hboxtosizeof#1#2{\setbox\SlashedBox=\hbox{#1} 
\hbox to 1\wd\SlashedBox{#2}} 

\def\ms#1{\setbox\SlashedBox=\hbox{$#1$}
\hbox to 0pt{\hbox to 1\wd\SlashedBox{\hfil/\hfil}\hss}#1}

\def\Z {\mbox{\sf{Z}} \hspace{-1.6mm} \mbox{\sf{Z}} \hspace{0.4mm}}

\newcommand{\ie}{{\em i.e.~}}
\newcommand{\eg}{{\em e.g.~}}

\newcommand{\be}{\begin{equation}}
\newcommand{\ee}{\end{equation}}
\newcommand{\ba}{\begin{eqnarray}}
\newcommand{\ea}{\end{eqnarray}}

\begin{document} 

\thispagestyle{empty}

\begin{flushright}
ROM2F/00/13
\end{flushright}

\vspace{1.5cm}

\begin{center}

{\LARGE {\bf Anomalies \& Tadpoles}} \\
\vspace{1cm} {\large Massimo Bianchi and Jose F. Morales}\\ 
\vspace{0.6cm} 
{\large {\it Dipartimento di Fisica, \ Universit{\`a} di Roma \  
``Tor Vergata''}} \\  {\large {\it I.N.F.N.\ -\ Sezione di Roma \ 
``Tor Vergata''}} \\ {\large {\it Via della Ricerca  Scientifica, 1}} 
\\ {\large {\it 00173 \ Roma, \ ITALY}} \\

\end{center}

\vspace{1cm}

\begin{abstract}

We show that massless RR tadpoles in vacuum configurations with 
open and unoriented strings are always related to anomalies. 
RR tadpoles arising from sectors of the internal SCFT 
with non-vanishing Witten index are in 
one-to-one correspondence with conventional irreducible anomalies.   
The anomalous content of the remaining RR tadpoles can be 
disclosed by considering 
anomalous amplitudes with higher numbers of external legs. 
We then provide an explicit parametrization 
of the anomaly polynomial in terms of the boundary
reflection coefficients, \ie one-point functions of massless RR fields
on the disk. After factorization of the reducible anomaly, 
we extract the relevant WZ couplings in the effective lagrangians.  

\end{abstract}

\vspace{4cm}
\noindent
\rule{6.5cm}{0.4pt} 

\newpage 

\setcounter{page}{1}

\section{Introduction}

Anomaly and tadpole vanishing conditions have been recognized for long as 
intriguingly related constraints in the construction of consistent vacuum 
configurations with open and unoriented strings. Indeed infinity and 
anomaly cancellations go hand by hand in the $SO(32)$ type I theory
\cite{gs}. Although considerable
experimental evidence has been accumulated in favour of a one-to-one 
correspondence between the two kinds of consistency conditions in models  
with ${\cal N}=(1,0)$ supersymmetry in $D=6$,
already in  $D=4$ there are exceptions to this rule 
and very little is known so far in the non-supersymmetric
case. At first sight it is not obvious why the vanishing of RR tadpoles, 
a transverse-channel constraint on the closed-string coupling to
boundaries and crosscaps, should in general be related to the vanishing 
of irreducible anomalies, a constraint on the spectrum that is naturally encoded 
in the direct-channel amplitudes.

The issue has been first addressed in \cite{pc} where the RR tadpole 
in $D=10$ was shown to induce a violation of BRST invariance on the 
string worldsheet. In \cite{msa,as} tadpole cancellation was proposed as the 
open-string analogue of closed-string modular invariance, that cuts off 
the UV region responsible of anomalies. Additional evidence for a
tadpole-anomaly correspondence was given in \cite{bs}, where some
${\cal N}=(1,0)$ supersymmetric models in $D=6$ and some
non-supersymmetric models in $D=10$ were shown to be  
free of irreducible anomalies thanks to the vanishing of RR tadpoles. 
The cancellation of the left-over reducible anomalies
relied on the presence of extra RR antisymmetric tensors that 
participate in a generalized Green-Schwarz mechanism \cite{as}.
The general belief on an IR-UV correspondence between RR tadpoles and 
anomalies has been reinforced by the advent of D-branes \cite{cjp}.
RR charge neutrality of vacuum configurations with closed unoriented 
strings requires the introduction of D-branes and their open-string 
excitations. The new trend motivated the study of \cite{bi}, where the
correspondence was analyzed for supersymmetric D-brane configurations 
at orbifold singularities in non-compact spaces. 
RR tadpoles from twisted sectors were identified with irreducible
gauge anomalies. 
More recently, an important contribution has been
given in \cite{ibanez}, where a careful comparison 
has been drawn in the context of geometric 
(supersymmetric) orbifold compactifications to $D=4,6$. 
In all cases studied in \cite{ibanez}, RR tadpoles 
associated to string amplitudes in sectors where the orbifold group 
acts without fixed tori have been put in a one-to-one correspondence
with irreducible anomalies. The remaining tadpole conditions
have been left as seemingly unrelated to the conditions for anomaly 
cancellation.  
The intrisic reasons for such a tadpole-anomaly correspondence 
and the range of its validity inside the more general web of open string 
compactifications remains unclear.  

The increasing interest in the phenomenological perspectives
of open string vacuum configurations with various patterns 
of supersymmetry breaking deserves a thorough analysis of this
long-standing problem. 
In the present paper we address the problem in full generality.
Our analysis relies on the topological properties of
the internal SCFT (super-conformal field theory) and encompasses not only 
unoriented descendants of geometric orbifolds and Gepner models,
but also descendants of asymmetric orbifolds and 
free fermionic constructions that result in left-right symmetric parent
theories.
We find that RR-tadpole conditions always correspond 
to the cancellation of some sort of anomaly in the effective theory.
The contribution to the anomalies from a given sector of the internal SCFT
is measured by its ``chiral'' Witten index, ${\cal I} = tr_{R} (-)^{F}$
\cite{ew}. 
This index is defined by a chiral vacuum amplitude in the odd spin 
structure, where the worldsheet supercurrent is periodic, 
and effectively counts the chiral asymmetry of each sector
\footnote{For definitions and applications of the index in the open
string context see \cite{ms}}.   
Tadpoles associated to sectors with non-vanishing Witten index
will be precisely identified with 
irreducible terms in the canonical anomaly polynomial.
This extends the notion of sectors without fixed tori that appear in 
the orbifold constructions discussed in \cite{ibanez}
to the present more general context. 
In addition an attentive study of the tadpoles that arise from
sectors with vanishing Witten index suggests their connection
to anomalous amplitudes involving a larger number of external 
insertions.  
 
We would like to stress that nowhere in the paper we will assume that 
the theory enjoys target space supersymmetry. Our discussion is 
completely general and applies to non-supersymmetric models that can 
arise for instance from superstring compactifications  
with or without brane supersymmetry. 

In order to keep the notation as compact as possible, the language we use is
the one of rational superconformal field theories but with obvious 
modifications one can accomodate in this setting all 
``irrational'' models studied so far such as tori and orbifolds thereof.
The only crucial ingredient in our considerations is the worldsheet 
consistency between direct (loop) and transverse (tree) 
channel for the relevant 
amplitudes. This imposes highly non-trivial constraints in the
construction of a sensible open string model.
Throughout the paper we will assume that a solution to these 
fundamental constraints (not to be confused with the tadpole-anomaly
conditions under consideration) has been found.   
Another important ingredient is the ``modular invariance'' of 
string amplitudes in the odd-spin structure which allows us to 
relate irreducible anomalies in the direct channel to RR tadpoles in the 
transverse channel.

After showing the equivalence between RR tadpoles  
and irreducible terms in the anomaly 
polynomial, we will pass to the study of the reducible ones.
We parametrize the anomaly polynomials in terms 
of reflection coefficients for the massless closed string 
states in front of crosscaps and boundaries.
We will find that, barring some minor ambiguities in $D=8,10$ if the 
Chan-Paton group is non semi-simple, the 
anomaly polynomial admits a unique factorization in terms of a sum of 
as many products as sectors that flow in the transverse channel and 
have non-vanishing Witten index. 
This allow us to factorize the reducible anomaly in such a way as
to immediately expose the R-R fields that participate in the 
generalized GSS mechanism and to extract the corresponding terms in the 
effective lagrangian. In particular one can immediately identify the 
``anomalous'' $U(1)$'s.

The plan of the paper is as follows. In Section 2 we briefly review the 
construction of open string descendants. In Section 3 we establish the 
precise relation between RR tadpoles for sectors with non-vanishing 
Witten index and irreducible anomalies.  
In Section 4 we study the
reducible part of the anomaly polynomials in any even dimensions
and extract the WZ couplings in the effective theories.
In Section 5 we discuss RR tadpoles arising from sectors 
with vanishing Witten index and elaborate on their connection to anomalous
amplitudes with higher number of external legs. 
In Section 6 we discuss our results and add some concluding remarks.

\section{Open string descendants }

In this section we review some relevant features in the
construction of open string descendants of generic 
left-right symmetric ``compactifications'' of type II 
and type 0 superstrings. 
By superstring compactification we mean any vacuum configuration 
whose worldsheet dynamics is governed by a superconformal field theory    
(SCFT) irrespective of the presence of unbroken target-space supersymmetry.

A closed string compactification to $D$ dimensions is defined 
by tensoring an internal SCFT, with
left and right moving central charges 
$(c,\bar{c})=({3 \over 2} (10-D),{3 \over 2} (10-D))$, with
a $(c,\bar{c})=({3 \over 2} D,{3 \over 2} D)$ spacetime
theory realized in terms of free worldsheet bosons and fermions 
$\{X^{\mu}, \psi^{\mu},\tilde{\psi}^{\mu}\}$. 
The Hilbert space of string states can be decomposed into a 
(generically infinite) sum ${\cal H} = \oplus_{i,j} {\cal H}_i \otimes
\bar{\cal H}_j$. The index $i$ labels the primary fields $\Phi^i$
of some chiral algebra ${\cal G}$, 
that includes an ${\cal N}=1$ superconformal algebra at least.
The left-moving subspace ${\cal H}_i$ thus consists in the tower of 
descendants under ${\cal G}$ of the groundstate $|h_{i}\rangle = 
\Phi^i|0\rangle $. 
The spectrum of the left-moving Hamiltonian $L_{0}$ in the sector ${\cal H}_i$ 
is conveniently assembled in the holomorphic character
\be
{\cal X}_{i}(\tau) = {\rm Tr}_{{\cal H}_i}^{\prime}\, q^{L_{0}-{c\over 
24}} \quad , 
\label{character}
\ee       
with $q=e^{2\pi i \tau}$. A prime in (\ref{character}) indicates the 
omission of the contribution of the bosonic zero-modes to the trace.
Moreover, it is always understood that states in the NS (sub)sector, 
where the worldsheet supercurrent is anti-periodic, enter with a 
plus sign and states in the R (sub)sector,
where the worldsheet supercurrent is periodic, enter with a minus sign.
This follows from modular invariance at two loops and implements the 
correct relation between spin and statistics. 
For the right-moving
excitations on the string worldsheet we assume $\bar{\cal G}\sim{\cal G}$ 
and define anti-holomorphic 
characters $\bar{{\cal X}}_{j}(\bar{\tau})$ similarly. 

The Lorentz transformation properties of states in ${\cal H}_i$ 
are encoded in the spacetime part of the character ${\cal X}_{i}$,
that can be decomposed into ``characters'' of the $SO(D-2)$ little group 
\ba
\chi_O+\chi_V &=&\left(\vartheta{0\brack 0} \over \eta^3\right)^{D-2\over2} 
\qquad
\chi_O-\chi_V = \left(\vartheta{0\brack {1\over 2}} \over 
\eta^3\right)^{D-2\over2}\nonumber\\
\chi_S+\chi_C &=&\left(\vartheta{{1\over 2}\brack 0} \over 
\eta^3\right)^{D-2\over2}\qquad
\chi_S-\chi_C =\left(-i\vartheta{{1\over 2}\brack {1\over 2}} \over 
\eta^3 \right)^{D-2\over2} \quad .
\label{lorentz}
\ea
The labels $O,V,S,C$ denotes the scalar, vector, spinor 
left (L) and spinor right (R) representation of the affine transverse Lorentz 
algebra at level $\kappa =1$. 
Once the contributions of the spacetime bosonic and fermionic coordinates 
are included, modular invariance ensures that any holomorphic character 
is given by a sum over spin structures:
\be
{\cal X}_{i} = \sum_{\alpha,\beta = 0, {1\over 2}} 
\left(\vartheta{\alpha\brack\beta} \over 
\eta^3 \right)^{D-2\over2} {\cal I}_{i}{\alpha\brack\beta} 
 = {\cal I}_{i}\Theta_{D} + \ldots \quad .
\label{chis}
\ee
and similarly for the antiholomorphic part.
In the second equality we have isolated the contribution 
of the odd spin structure on which our analys will be focussed.
In this spin-structure worldsheet bosons and fermions 
share the same boundary conditions. The worldsheet 
supercurrent is periodic and the 
contributions of the massive excitations to the sum
cancel against each other. The contribution of the spacetime super-coordinates, 
\be
\Theta_{D} = \left(-i\vartheta{{1\over 2}\brack {1\over 2}} \over 
\eta^3 \right)^{D-2\over2} \quad ,
\ee
though formally zero, is a convenient book-keeping of the chiral 
asymmetry. The result of the trace in each sector of
the internal SCFT theory,  
${\cal I}_{i} \equiv {\cal I}_{i}{{1\over 2}\brack {1\over 2}}$,
is an integer (the Witten index \cite{ew}) that 
counts the difference between the number of bosonic and fermionic 
ground-states. From the right-hand side of 
(\ref{chis}) we notice that ${\cal I}_{i}$ effectively counts the difference
between the number of left- ($L$) and right- ($R$) ``handed'' spacetime fermions. 

The spectrum of perturbative closed string states
is then packaged in the one-loop (torus) partition function. 
Neglecting overall volume factors
and the modular integration ($\int_{{\cal F}}\frac{d^2 \tau}{\tau_2^2}$) 
one finds 
\be
{\cal T} = \tau_{2}^{-{D-2\over 2}}
\sum_{ij} {\cal T}_{ij} {\cal X}_{i} \bar{\cal X}_{j} \quad .
\label{torus}
\ee
${\cal T}_{ij}$ are positive integer coefficients with 
${\cal T}_{00}=1$ for the uniqueness of the identity sector, \ie of the graviton. 
The powers of $\tau_2$ arise from the momentum integrations in the 
non-compact directions.
The condition for left-right symmetry on the worldsheet translates into the 
constraints ${\cal T}_{ij} = {\cal T}_{ji}$. The coefficients 
${\cal T}_{ij}$ are also 
highly restricted by one-loop modular invariance. The characters are 
known
to provide a unitary representation of the modular group $SL(2,\Z )$ 
generated by the transformations $T$ and $S$ under which 
\ba
T: \qquad {\cal X}_i(\tau+1) &=& e^{2\pi i(h_{i}-c/24)}{\cal X}_i(\tau)
\nonumber\\
S: \qquad {\cal X}_i \left(-{1 \over\tau}\right) &=& \sum_{j}
(i \tau)^{-\frac{D-2}{2}}S_{ij}{\cal X}_j(\tau) \quad .
\label{modular}
\ea
After the resolution of fixed-point ambiguities, related to the presence of 
different sectors with the same character, the transformation $S$ is 
represented by a symmetric matrix that satisfies
$(ST)^{3} = S^{2} = C$, with $C$ the charge conjugation matrix.

In most of our subsequent discussion, we will restrict our 
attention 
to the case in which the boundary conditions associated to the 
introduction of open and unoriented strings preserve 
the diagonal combination of the worldsheet symmetries of the bulk theory
together with their target-space byproducts. This generalizes the 
notion of a BPS D-brane.
With a slight change of notation, that amounts to decomposing the 
characters of the parent theory into their irreducible components with respect 
to some lower (super-)symmetry, one can easily accomodate models with    
brane (super-)symmetry breaking. This generalizes the 
notion of a non-BPS D-brane. 
Moreover we will display formulas which are more akin
to the rational context, where the number of characters is
finite. With some additional care, the results can be adapted to  
irrational contexts such as toroidal or orbifold compactifications
at irrational values of the moduli (radii, shapes and Wilson lines).

At generic points of the moduli space of toroidal compactifications \cite{bps} 
the index $i$ may be thought to run 
over an infinite number of primary fields, one for each independent 
choice of internal momenta and windings. 
Similarly in orbifold compactifications, only states that
are invariant under the action of the orbifold group will
enter the trace (\ref{character}).  
Chiral string excitations are then organized according to their eigenvalues 
under the action of the orbifold group. Taking $\Z_{N}$ for simplicity, 
one is lead 
to define the characters \cite{bmp}
\be
{\cal X}_{gh}=\frac{1}{N} \sum_{k=0}^{N-1}\, \rho_{g k} \, \omega_{h}^{k}
\label{chorb}
\ee
with $g,h=0,1,\ldots N-1$ labelling all possible twists in the $\sigma$
and $\tau$ directions and $\omega_{h}=e^{2\pi i h\over N}$.
In the unwisted sectors and in sectors with fixed tori,
there are infinitely many characters 
depending on the choices of allowed momenta and windings. In 
twisted sectors without fixed tori, 
a finite number of twisted characters lives at each fixed point. 
The chiral amplitudes $\rho_{gh}$ are defined by the traces
\be
\rho_{gh}={\rm Tr}_{g}\, h \,q^{L_{0}-c/24}.
\ee
in the $g$-twisted sector. Explicit expressions for these traces 
in $D=4,6$ can be found in the appendix A of \cite{bmp}. 

By suitably tuning the parameters, any toroidal or orbifold 
compactification can be described in terms of a rational SCFT. Moreover in both
the rational and the irrational contexts the number of 
massless characters, \ie sectors with massless ground-states, is finite. 
Since only these states will
be relevant to our analysis the two cases can be treated in parallel.
Focussing on the odd spin-structure,
where anomalies potentially reside,
one is effectively dealing with the ``topological'' part of 
the theory that is largely independent of the moduli of the SCFT.
For instance, only massless sectors with 
non-vanishing Witten index enter the computation of the Euler number of the 
``compactification manifold'' ${M}$ \cite{ew,eoty},
\be
\chi({M}) = \sum_{i,j}{\cal T}_{ij}{\cal I}_{i}{\cal I}_{j} 
\quad ,
\ee
and the other curvature invariants that allow one to identify
the topological class the vacuum configuration belongs to.
On the other hand, massive string states do not contribute to the Witten index
since they always come in Bose-Fermi degenerate pairs.
Ground-states that can become massless at special points 
(\eg decompactification limit) of the 
moduli space, though not contributing to the na\"ive 
Witten index, enter the computation of anomalous amplitudes in a subtle 
way.

Let us now consider the unoriented descendant of a left-right 
symmetric closed-string theory, the ``parent'' theory, 
defined by (\ref{torus}). For definiteness we will consider 
orientifold reductions by the worldsheet parity operator
$\Omega$. More general unoriented descendants that
combine $\Omega$ with other internal symmetries can be
similarly studied. The unoriented closed string spectrum
is determined by halving (\ref{torus}) and adding to it the Klein-bottle 
projection.
Up to overall volume factors and the modular integration 
($\int\frac{dt}{t}$) one has 
\be
{\cal K} = {1\over 2} {\rm Tr}_{closed}\, ( \Omega \, q^{L_{0}-{c\over 
24}} ) ={1\over 2\,t^{D\over 2}} 
\sum_{i} K^{i}{\cal X}_{i}(2it) 
= {1\over 2\,t^{D\over 2}} 
\sum_{i} K^{i}{\cal I}_{i}
\Theta_{D}(2it) + \ldots \quad .
\label{klein}
\ee
In the right hand side, for later convenience, only the contribution of the 
odd spin-structure $\Theta_{D}$ has been displayed.
The sum in (\ref{klein}) is restricted to   
the diagonal terms (${\cal T}_{ii} \neq 0$) in (\ref{torus}).  
For permutation modular invariants (${\cal T}_{ij} = 0,1$) 
the Klein-bottle 
coefficients $K^{i}$, implementing the action of $\Omega$ on 
${\cal H}_i \otimes \bar{\cal H}_i$,
are related to the ones in the torus partition 
function through $K^{i}= \pm {\cal T}_{ii}$.

Whenever the action of $\Omega$ is not free
the consistency of the theory requires the inclusion 
of boundaries and the corresponding open string sectors 
ending on them. The open string partition function consists in the 
Annulus and M\"obius-strip amplitudes. Neglecting the overall volume 
factors and the modular integration ($\int \frac{dt}{t}$) one has     
\ba
{\cal A} &=& {1\over 2\, t^{D\over 2}} \sum_{i,a,b} A^{i}_{ab} n^{a} n^{b} 
{\cal X}_{i} \left( {it\over 2}\right)  
={1\over 2 \,t^{D\over 2}}
\sum_{i,a,b} A^{i}_{ab} n^{a} n^{b}{\cal I}_{i}
\Theta_{D}\left( {it\over 2}\right) + \ldots \quad .
\nonumber\\
{\cal M}&=&{1\over 2\,t^{D\over 2}} 
\sum_{i,a} M^{i}_{a} n^{a}  \widehat{\cal X}_{i}
\left( {it\over 2} + {1\over 2}\right)  
={1\over 2\,t^{D\over 2}}
 \sum_{i,a} {\cal I}_{i} M^{i}_{a} n^{a}
\widehat\Theta_{D}\left( {it\over 2} + {1\over 2}\right)
\label{direct} \quad
\ea
A proper basis of real hatted characters $\widehat{\cal X}_{i} =
T^{-{1/2}} {\cal X}_{i}$ has been introduced in ${\cal M}$ \cite{bs}.  
The indices $a,b$ run over the number of boundaries,
\ie independent Chan-Paton charges, with integer multiplicities $n^a$.
Barring some exceptions to be discussed later,
the number of independent charges one should introduce 
equals the number of characters 
${\cal X}_{i}$ that are paired with their charge 
conjugates $\bar{\cal X}_{i^{C}}$ in the parent theory (\ref{torus}).
The integers $A^{i}_{ab}$ and $M^{i}_{a}$ 
count the number of times the sector $i$ runs in 
an Annulus and M\"obius-strip loop, respectively, of  
open strings with ends on $(a,b)$ and $(a,a)$ respectively.
Here and in the following, we will not 
distinguish between ``complex'' (unitary) and ``real'' 
(orthogonal or symplectic) Chan-Paton multiplicities $n^a$. 
The sum over $a$ will include 
two contributions for the former and only one for the latter, thus 
reproducing not only the correct dimensions of the representations but 
also the correct orientation of the boundary.

The quantities appearing in (\ref{klein}), (\ref{direct}) are 
highly constrained by two consistency requirements. 

The first is the requirement of a consistent interpretation of
the transverse channel of these amplitudes as the tree-level exchange 
of closed string states between
boundaries and/or crosscaps. This implies that
under $t\rightarrow 1/t$ the coefficients of the above three amplitudes should
reconstruct perfect squares   
\ba
\widetilde{\cal K} &=& {2^{D\over 2}\over 2} 
\sum_{i} (\Gamma^{i})^{2} {\cal X}_{i}(q) 
= {2^{D\over 2}\over 2} 
\sum_{i} (\Gamma_{i})^{2}{\cal I}_{i}
\Theta_{D}(q) + \ldots \nonumber\\
\widetilde{\cal A} &=& {2^{-{D\over 2}}\over 2} 
\sum_{i,a} (B^{i}_{a}n^{a})^{2}{\cal X} _{i}(q) 
= {2^{-{D\over 2}}\over 2} 
\sum_{i,a} (B^{i}_{a}n^{a})^{2}{\cal I}_{i}
\Theta_{D}(q) + \ldots \nonumber\\
\widetilde{\cal M} &=& {2\over 2} 
\sum_{i,a} ( \Gamma^{i} B^{i}_{a}n^{a}) \widehat{\cal X}_{i}(-q) 
=-{2\over 2} 
\sum_{i,a}(\Gamma^{i} B^i_a n^a) {\cal I}_{i}
\widehat\Theta_{D}(-q) + \ldots
\label{transverse} 
\ea
The relative powers of 2 result from the different rescalings of the modular 
parameters ($\tau_{{\cal K}}= 2it, \tau_{{\cal A}} = it/2, 
\tau_{{\cal M}} = it/2 + 1/2$) 
that naturally enter the definition of the amplitudes in 
the direct channel. These rescalings are necessary in order for the 
amplitudes in the transverse channel to be expressed in terms of the common 
length of the tube $\ell = - {1\over 2\pi} \log {q}$.
The coefficients $\Gamma^{i}$ and $B^{i}_{a}$ should then be
interpreted as the reflection coefficient (one-point function) 
on a crosscap and on a boundary of type $a$ respectively.
They are related to the integer coefficients in the direct channel 
by suitable modular transformations:
\ba
K^{i} &=& \sum_{j} S^{i}{}_{j}\Gamma^{j}\Gamma^{j} \nonumber\\
A^{i}_{ab} &=& \sum_{j} S^{i}{}_{j}B^{j}_{a}B^{j}_{b} \nonumber\\
M^{i}_{a} &=& \sum_{j} P^{i}{}_{j}\Gamma^{j} B^{j}_{a}
\label{worldsheet}
\ea
with $S$ defined in (\ref{modular}) and $P\equiv T^{1/2} S T^2 S T^{1/2}$.  

In addition one should require that fluxes of 
massless RR-fields should not be trapped inside the ``compactification'' 
manifold. This leads to the tadpole 
cancellation conditions:
\be
 2^{D/2}\Gamma^{i} + \sum B^{i}_{a} n^{a} = 0
 \label{tadpolec}  
\ee
where $i$ runs over any sector that flow in the transverse channel and
contain RR massless states.
Understanding how to precisely relate these conditions to another consistency
requirement, the absence of irreducible anomalies in the low-energy field
theory, will be the main subject of our investigation.

Finding a solution to (\ref{worldsheet}) is in general a highly non-trivial
step in the construction of an open descendant. We will always assume 
that this has been properly done. 
It is worth stressing that a ``canonical'' solution always exists
if the torus modular invariant is given by the charge 
conjugation matrix ${\cal T}_{ij} = C_{ij}$ (``geometric 
compactifications'' such as tori, symmetric 
orbifolds\footnote{As we will see, in
some orbifold models this may require some identifications among the 
Chan-Paton charges.} and Gepner 
models belong in this category).
In this case the number of independent Chan-Paton 
charges is exactly equal to the number of characters and 
one can use the same kind of indices, say $i,j,\ldots$, to label both.
The ansatz \cite{cardy,bs,fpss} amounts to taking 
\ba
K^{i}&=& Y_{00}{}^{i} \nonumber\\  
A_{ij}{}^{k} &=& N_{ij}{}^{k} \nonumber\\
M_{j}{}^{i} &=&Y_{0j}{}^{i} \quad ,
\ea
where 
\be
N_{ij}{}^{k}=\sum_{l} {S_{il}S_{jl}(S^{\dagger})^{lk} \over S_{0l}}
\ee
are the fusion rule coefficients, that 
compute how many independent couplings of the primary fields $i$ and 
$j$ give the primary field $k$, and 
\be
Y_{ij}{}^{k} = \sum_{l} {S_{il}P_{jl}(P^{\dagger})^{lk} \over S_{0l}}
\ee
are (not necessarily positive) integers that satisfy 
$Y_{ij}{}^{k}=N_{ij}{}^{k}(mod \: 2)$. 
The boundary and crosscap reflection coefficients then read
\ba
\Gamma^{i}&=&{P^{i}{}_{0} \over \sqrt{S_{0i}}}\nonumber\\
B^{i}{}_{j}&=&{S^{i}{}_{j} \over \sqrt{S_{0i}}} \quad .
\ea

Notice that quantities in the transverse
amplitudes (\ref{transverse}) are related to the spectra
(\ref{klein}), (\ref{direct}) through the $S$ and $P$ modular
transformations, whose
details vary from model to model. Fortunately this is not the
case for the contribution of the odd-spin structure 
in (\ref{chis}). Each sector of the internal SCFT only enters 
through its Witten index ${\cal I}_{i}$ that is modular invariant and the 
modular transformations of $\Theta_{D}$ are simply
\ba
\Theta_{D}(-1/it)&=& (it)^{D-2\over 2} \Theta_{D}(it)\nonumber\\
\Theta_{D} (it+1) &=& \Theta_{D} (it)
\label{Sodd} \quad .
\ea
This elementary observation will be the basis of all our manipulations in what 
follows.

\section{Anomalies in open string descendants} 

In this section we discuss the relation between RR tadpoles
and irreducible anomalies in a generic even-dimensional open descendant.
We first review (and slightly adapt) the results of 
\cite{ikk,ss}, which allow us to reproduce the
anomaly polynomial by an string computation 
in the unoriented descendant.  
Our strategy will be to use modular invariance of the odd-spin structure
(\ref{Sodd}) to parametrize the anomaly polynomial in terms of the 
reflection coefficients in the transverse channel. 
On the one hand, this allows us to identify the precise 
combination of RR tadpoles
that corresponds to the irreducible gravitational anomaly.
On the other hand, using the completeness properties of the 
boundary reflection coefficients, this will enable us to map 
RR tadpoles in sectors with non-vanishing Witten index 
to irreducible gauge anomalies. After imposing the tadpole/anomaly 
conditions, the anomaly polynomial turns out to be reducible. It thus
admits a simple factorization that suggests the participation of 
various RR antisymmetric tensors of different rank to a
generalized mechanism of anomaly cancellation. 
  
Anomalies are encoded in the odd-spin structure part
of the one-loop string amplitudes
\be
\langle \prod_{f=1}^{N} V_{F}(p_{f}, \xi_{f}) 
\prod_{g=1}^{M} V_{G}(p_{g}, h_{g}) \rangle
\ee
involving $N$ gauge field $V_{F}$ and $M$ graviton $V_{G}$ 
vertex operators with polarization and momenta 
$\xi_{f},p_f$ and $h_{g},p_g$, respectively. The number of external 
legs is such that
$N+M= D/2 + 1$, where $D$ is the (even) number of non-compact dimensions. 
One of the vertices has to be taken
with a longitudinal polarization.
Since the parent theory is assumed to be anomaly free
\footnote{Modular invariance ensures that this is always the case.} 
anomalous contributions to the above amplitudes only arise from the  
Klein bottle, Annulus and M\"obius strip. 
In the odd spin-structure there is one supermodulus (zero-mode of the
spin 3/2 bosonic superghost $\beta$) and one conformal Killing spinor
(zero-mode of the spin -1/2 bosonic superghost $\gamma$).
In order to dispose of the former a picture changing operator is to 
be inserted. In order to dispose of the latter, one 
of the vertices (let's say the one with longitudinal polarization) 
should be taken in the  $(-1)$ picture. As a result the total superghost 
charge remains zero as expected for surfaces with vanishing Euler 
characteristic. 
After simple manipulations (see \cite{ss} for details) 
a generating function for the anomalous amplitudes can be written as 
\be
A = \int_{0}^{\infty} dt {d\over dt} \langle e^{-S_{0} + S_{F} + S_{R}}
\rangle_{odd}
\ee
where $S_{0}$ is the free action, and the exponentiated 
effective vertex operators $S_{F},S_{R}$ are given by 
\ba
S_{F} &=& \oint ds \,F^{a} \lambda^{a} \nonumber \\
S_{G} &=& \int d^{2}z \,R_{\mu\nu} [ X^{\mu} (\partial + \bar\partial) 
X^{\nu} + (\psi^{\mu} - \tilde\psi^{\mu})  (\psi^{\nu} - 
\tilde\psi^{\nu}) ]
\ea
with
\be
F^{a} = {1\over 2} F^{a}_{\mu\nu}(\psi_{0}^{\mu} \psi_{0}^{\nu}) 
\qquad 
R_{\mu\nu} = {1\over 2} R_{\mu\nu\rho\sigma}(\psi_{0}^{\rho} 
\psi_{0}^{\sigma}) \quad .
\ee
bilinears in the zero modes of non-compact fermionic coordinates. 
After integration over the grassmanian variables, $F^{a}$ and 
$R_{\mu\nu}$ behave as two-forms. 

The above determinant has been computed in \cite{mss} (see also 
\cite{ss1} for additional insights). The crucial point 
is that the determinant, being defined by a trace in the odd-spin
structure where worldsheet bosons and fermions share the same boundary 
conditions, is given by a $t$-independent topological invariant.   
Each sector of the internal SCFT only enters through its
Witten index ${\cal I}_{i}$ and the final result can be written 
as \cite{mss}
\ba
{\cal K}_{odd} &=&
      -{1 \over 2}\sum_{i} {\cal I}_{i} K^{i}\,I_{A}(R)
\label{Kanomaly}\nonumber\\
{\cal A}_{odd} &=& 
 ={1 \over 4}\sum_{i,a,b}{\cal I}_{i} A^{i}_{ab} ch_{n^{a}}(F) ch_{n^{b}}(F)
 I_{{1/2}}(R)
            \nonumber\\
{\cal M}_{odd} &=& 
    {1 \over 4}\sum_{i,a}{\cal I}_{i} M^{i}_{a} n^{a}ch_{n^{a}}(2F) 
          I_{{1/2}}(R) 
            \label{Manomaly}
            \ea
where $ch(F)$ is the Chern character and $I_{A}(R)$ and $I_{{1/2}}(R) $ 
represent the contributions to the gravitational anomaly 
of a self-dual antisymmetric tensor
and a complex spin 1/2 L-fermion respectively.
The additional factor of one-half in the Annulus and M\"obius-strip 
amplitudes reflects the fact that they are counting real fermions. The 
relation between spin and statistics is responsabile for the extra minus 
sign of the Klein-bottle contribution with respect to the Annulus and 
M\"obius strip. The former can only contribute loops of 
(anti)self-dual antisymmetric tensors while the latter can only contribute 
fermionic loops. 
In terms of the Hirzebruch polynomial $\widehat{\cal L}(R)$ and A-roof genus 
$\widehat{\cal A}(R)$ one has
\ba
I_{A}(R)&=& {1\over 8}\widehat{\cal L} (R) =
  -{1\over 8} + {1\over 48} R^{2} 
+{1\over 32} \left( {7\over 45} R^{4} - {1\over 9} (R^{2})^{2} 
\right)\nonumber\\
&&+{1\over 128} \left( {496\over 2835} R^{6} - {588 \over 2835} R^{2} R^{4} 
+{140\over 2835} (R^{2})^{3} \right) + \ldots
 \\
I_{{1/2}}(R) &=& \widehat{\cal A}(R)= 1 + {1\over 48} R^{2} 
+{1\over 32} \left( {1\over 180} R^{4} + {1\over 72} (R^{2})^{2} \right)
 \nonumber\\
 &&+{1\over 128} \left( {1\over 2835} R^{6} + {1\over 1080} R^{2} R^{4} 
 +{1\over 1296} (R^{2})^{3} \right) + \ldots
 \\
 ch_{n^a}(F) &=& tr_{n^a} (\exp{iF}) 
\equiv\sum_{k} {1\over k!} F_a^{k}
 \ea
where by $R^{2m}$ we mean ${\rm tr}_{V} R^{2m}$ and wedge products are 
always understood. In the last line we have 
introduced the shorthand notation $F_a^{k}=tr_{n^a}\,F^{k}$ for later convenience.
 
The absence of anomalies in 
the parent theory allows one to 
rewrite the Klein-bottle contribution in terms of the spectrum of 
massless closed-string states of the unoriented descendant. 
As in \cite{mss}, we find
the expected field-theory result
 \be
 {\cal K} =-{1 \over 2} \sum_i {\cal I}_i K_i I_{A}(R) = 
 (n^{L}_{A}- n^{R}_{A}) I_{A}(R) + (n_{3/2}^{L}-n_{3/2}^{R})I_{3/2}(R) 
 +(n_{1/2}^{S}- n_{1/2}^{C})I_{1/2}(R) \quad ,
 \label{kmassless}
\ee
where $n^{L,R}_{A}$, $n_{1/2}^{L,R}$ and $n_{3/2}^{L,R}$ are respectively 
the numbers of
antisymmetric tensors, spin $1/2$ and spin $3/2$ massless closed-string states
with definite chirality properties and 
and $I_{{3/2}}(R) = (ch_{V}(R) - 1) \widehat{\cal A}(R)$. 
Similarly the contribution to the anomaly polynomial 
coming from the Annulus and M\"obius-strip amplitudes (\ref{Manomaly}) 
reproduces the expected field theory result.
Piecing everything together yields
\be 
{\cal P}(R,F) =\sum_{i} {\cal I}_{i} \left[
 - {1\over 2} K^{i}I_{A}(R) +{1\over 4} \left( \sum_{a,b} 
 A^{i}_{ab} ch_{n^{a}}(F) ch_{n^{b}}(F) + 
 \sum_{a} M^{i}_{a}ch_{n^{a}}(2F) \right)I_{{1\over 2}}(R)\right] 
 \quad .
\label{dirtot} 
\ee
In (\ref{dirtot}) the trace in the (anti-)symmetric representation 
of the gauge group $a$ is written  
in terms of the trace in the $n_{a}$-dimensional 
fundamental representation by means of
\be
ch_{{1\over 2}n^{a}(n^{a}\pm 1)} (F) =
{1\over 2} \left[ ch_{n^{a}}(F)^2\pm ch_{n^{a}}(2F) \right]  \quad .
\label{anti-sym} 
\ee

Few remarks are in order.
First, modular invariance of the parent closed string theory not only 
implies the vanishing of the irreducible anomalies but also of 
reducible ones, \ie no GS mechanism is at work in the parent theory.
This is well-known for left-right symmetric compactifications but 
we suspect it to be true in any perturbative vacuum configuration.
Second, the Klein bottle only contributes to the pure gravitational 
anomaly. We have not bother to turn on the field-strengths of any 
closed-string vector boson since all massless fermions of a left-right symmetric 
closed-string theory are neutral with respect to them. 
It is well-known that RR vector bosons are not minimally 
couple to perturbative string states and that non-abelian NS-NS vector 
bosons may be present in left-right symmetric theories only if 
not supersymmetric. 
Finally the CPT theorem in $D=4k$ 
dimensions requires $n^{L}_{A}= n^{R}_{A}$, $n_{3/2}^{L}= n_{3/2}^{R}$
and $n_{1/2}^{L} = n_{1/2}^{R}$. As a consequence closed string 
states do not 
contribute to the anomaly polynomial in these dimensions. The 
Klein bottle  
may give a non-trivial contribution only in $D=4k+2$ dimensions.
By the same token, the open string sector may contribute to the 
anomaly in $D=4k$ dimensions only when the massless fermions are 
chiral and belong to 
complex representations of the Chan-Paton group.
Since spinorial representations of orthogonal groups cannot appear 
perturbatively and symplectic groups only admit (pseudo-)real 
representations, only when unitary groups 
are present with associated ``complex'' 
Chan-Paton charges can the theory be chiral in $D=4k$ dimensions.
In $D=4k+2$ dimensions, on the contrary, barring few exceptions, any 
unpaired massless fermion contributes to gauge, gravitational and 
mixed anomalies.

We are now ready to pin down  
the UV-IR correspondence between anomalies and tadpoles in open 
descendants.  
Using the general relations (\ref{worldsheet}) and the fact that 
the Witten index, being a $t$-independent integer number is invariant
under modular transformations, \ie 
\ba
\sum_{i}{\cal I}_{i} S^{i}{}_{j} &=& {\cal I}_{j} \nonumber\\ 
\sum_{i}{\cal I}_{i} P^{i}{}_{j} &=& {\cal I}_{j} \label{wittinv} \quad ,
\ea
one can freely transfer the information encoded in the massless sectors with 
non-vanishing Witten index from one channel to the other.
Indeed, saturing (\ref{worldsheet}) with ${\cal I}_i$ and using 
(\ref{wittinv}) one can establish the much simpler dictionary
\ba
\sum_{i} {\cal I}_{i} K^{i} &=& \sum_{i} 
{\cal I}_{i}\Gamma^{i}\Gamma^{i}\nonumber\\
\sum_{i} {\cal I}_{i} A^{i}_{ab} &=& 
\sum_{i} {\cal I}_{i}B^{i}_{a}B^{i}_{b}\nonumber
\\
\sum_{i} {\cal I}_{i} M^{i}_{a} &=& 
\sum_{i} {\cal I}_{i}B^{i}_{a}\Gamma^{i} \quad ,
\ea
that does not explicitly involve any modular transformation.
Plugging these relations into the anomaly 
polynomial  (\ref{dirtot}) yields 
\be 
{\cal P}(R,F) = \sum_{i} {\cal I}_{i}
\left[ - {1\over 2}(\Gamma^{i})^{2} I_A(R) +
{1\over 4} \left( \sum_{a} B^{i}_{a}ch_{n^{a}}(F) \right)^2
I_{{1 \over 2}} (R)-
{1\over 4}\Gamma^{i}\sum_{a}B^{i}_{a} ch_{n^{a}}(2F) 
I_{{1 \over 2}} (R) \right] \quad .
\label{polinomial}
\ee
The consistency of the effective theory requires the cancellation 
of the irreducible terms in (\ref{polinomial}).
In particular, for the pure gravitational 
anomaly, using the relation 
\be
I_{A}^{D}(R)|_{irr} = \frac{2^{D/2}(2^{D/2}-1)}{2} I_{1/2}^{D}(R)|_{irr}
\label{irreducible} \quad ,
\ee
the vanishing of the coefficient of $tr_{V} R^{\frac{D}{2}+1}$
in the expansion of (\ref{dirtot}) requires
\be 
\sum_{i} {\cal I}_{i}
\left( 2^{D/2} \Gamma^{i} + \sum_{a}B^{i}_{a} {n^{a}} \right)  
\left((2^{D/2}-1)\Gamma^{i} + \sum_{a}B^{i}_{a} {n^{a}} \right) = 0 
\label{quasir} \quad .
\ee
Notice that this is a specific linear combination of the RR tadpole
conditions (\ref{tadpolec}). In particular the cancellation of the
tadpoles automatically implies the cancellation of this
irreducible anomaly.

Similarly the absence of irreducible gauge anomaly, \ie the terms 
$tr_{n^b} F^{\frac{D}{2}+1}$
in the expansion of (\ref{dirtot}), is equivalent to the conditions
\footnote{Notice that $U(n)$ field
strengths couple to $n$ and $\bar{n}$ complex charges, and therefore
the corresponding gauge anomalies (both irreducible and reducible)  
will always include the sum of two terms involving traces
in the $n$ and $\bar n$ after using (\ref{anti-sym}).}
\be 
\sum_{i} {\cal I}_{i}
\left( 2^{D/2} \Gamma^{i} + \sum_{a}B^{i}_{a} {n^{a}}\right) B^{i}_{b} = 0 
\label{quasi} \quad .
 \ee
As apparent one has in principle one tadpole condition for each 
factor in the Chan-Paton group. Since the number of independent 
boundary conditions coincides with the number of independent 
Chan-Paton multiplicities, one may suspect to be on the right track.
Indeed one can turn an index $a$, that labels the factors in the 
Chan-Paton 
group, into an index $i$, that labels the characters that flow in the 
transverse channel, by making use of the completeness relation, 
\cite{sas} 
\be
\sum_{a} B^{i}_{a}B^{j}_{a} = {\Pi^{{ij}}\over \sqrt{S_{0i}S_{0j}}}
\label{complete} \quad ,
\ee
valid for any permutation modular invariant parent theory.
$\Pi^{ij}$ is the projector onto the set of characters that flow
in the transverse channel and $S_{0i}$ are the elements in the first 
row/column of the matrix $S_{ij}$ representing modular 
$S$-transformations (\ref{modular}). $S_{0i}$ are 
always positive and are sometimes called ``quantum dimensions''.

By saturating the Chan-Paton index $b$ in (\ref{quasi})
with $B^{j}_{b}$ we are left with 
\be
{\cal I}_{i}\left(2^{D/2}\Gamma^{i}+B^{i}_{a} n^{a}\right)\Pi^{{ij}}=0
\label{quasitadpole} \quad .
\ee
This precisely reproduces all tadpole cancellation conditions
for massless R-R closed string states flowing in the
transverse channel (as the presence of $\Pi^{{ij}}$ indicates) 
and belonging to sectors with non-vanishing Witten index 
${\cal I}_{i}\neq 0$. It is amusing to observe that
in a consistent open descendant in $D=4k+2$ 
the absence of irreducible gauge anomalies
always implies the absence of irreducible gravitational
anomalies. The vanishing of the first factor in (\ref{tadpolec}) is 
selected as the relevant tadpole condition. 

A comment is in order. As the attentive reader might have observed,
a rephrasing of the above conclusion
is necessary in the context of orbifolds.
Indeed, in these constructions the worldsheet consistency 
(\ref{worldsheet}) between the direct (one-loop) and
transverse (tree) channel often restricts the number of allowed CP charges.
As a result the index $a$ often runs over a smaller subset than 
the set of characters flowing in the transverse channel. 
If this is the case, the index $i$ in the completeness
conditions (\ref{complete}) and tadpoles (\ref{quasir},\ref{quasi})
should be understood as running over an smaller subset defined by
the linear combination of characters flowing in this channel.
In the new basis (usually conveniently written in terms of
the chiral amplitudes) additional ``sectors'' with vanishing
Witten index arise and the corresponding tadpole
information is consequently lost in (\ref{quasi}).

Modulo this subtlety,
we conclude that any massless RR tadpole arising from a sector 
with non-vanishing
Witten index can be identified with some irreducible anomaly.  
In particular string theory defines a notion of irreducible gauge 
anomalies even for non-semisimple groups as the ones originating from 
amplitudes in which all the vertex operators are inserted on the same boundary.
Let us now concentrate on the left over
reducible part of the anomaly polynomial (\ref{polinomial}).

\section{Anomaly polynomial and WZ couplings}

Once the tadpole conditions are imposed the anomaly 
polynomial ${\cal P}(R,F)$  
is reducible and can be factorized into a sum of products.
This factorization allows one to extract the WZ and 
anomalous couplings of the R-R 
massless states that participate in a generalized GSS mechanism of 
anomaly cancellation \cite{as2}
\footnote{Explicit examples of WZ lagrangians in 
supersymmetric $D=4,6$ open string compactifications
have been worked it out recently in \cite{ss,rs}}.
Since the factorization is somewhat sensitive to the (even) dimension of the 
non-compact part of the target space it is convenient to decompose the 
total anomaly polynomial in terms of its $D$-dimensional components
\be
{\cal P}(R,F) = \sum_{D} {\cal P}_{D}(R,F) \quad ,
\label{pold}
\ee
where ${\cal P}_{D}(R,F)$, the anomaly polynomial in $D$-dimension,
is a $(D+2)$-form.

After some simple algebra one finds that the factorization is 
almost always unique. Only in $D=8$ and $D=10$ we find mild 
ambiguities. 
The anomaly polynomials in $D$ dimensions explicitly factorize as
\ba
{\cal P}_{0}(R,F) &=& 0
\nonumber\\
{\cal P}_{2}(R,F) &=&{1 \over 4} \sum_{i,a,b} {\cal I}_{i} B^{i}_{a} B^{i}_{b}
\left[X^{a}_{2} X^{b}_{2}\right]\nonumber\\
{\cal P}_{4}(R,F) &=&{1 \over 4} \sum_{i,a,b} {\cal I}_{i} B^{i}_{a} B^{i}_{b} 
\left[2 X^{a}_{2} X^{b}_{4}\right]
\nonumber\\
{\cal P}_{6}(R,F) &=& {1\over 4} \sum_{i,a,b} {\cal I}_{i} B^{i}_{a} B^{i}_{b} 
\left[ X^{a}_{4} X^{b}_{4}+2 X^{a}_{2} X^{b}_{6}(1)\right] 
\nonumber\\
{\cal P}_{8}(R,F) &=& {1\over 4} \sum_{i,a,b} {\cal I}_{i} B^{i}_{a} B^{i}_{b} 
\left[2 X^{a}_{4} X^{b}_{6}(\xi_{b}) +2 X^{a}_{2} X^{b}_{8}(\xi_{b}) \right]
\nonumber\\
{\cal P}_{10}(R,F) &=&{1\over 4} \sum_{i,a,b} {\cal I}_{i}  B^{i}_{a} B^{i}_{b}
\left[2 X^{a}_{2} X^{b}_{10}(\xi_{b}) +2 X^{a}_{4} X^{b}_{8}(1)
+ X^{a}_{6}(\xi_{a}) X^{b}_{6}(\xi_{b}) \right] \quad ,
\ea
where, in order to keep the above formulas as compact as possible, we have 
introduced the definitions
\ba
X^{a}_{2} &=& i{F_{a}}
\nonumber\\
X^{a}_{4} &=& {i^{2} \over 2!}({F^{2}_{a}} - {n^{a} \over 32} R^{2})
\nonumber\\
X^{a}_{6} (\xi_{a}) &=&{i^{3} \over 3!}({F^{3}_{a}} - 
{\xi_{a} \over 16} {F_{a}} R^{2}) 
\nonumber\\
X^{a}_{8} (\xi_{a}) &=&{i^{4} \over 4!}({F^{4}_{a}} +
{\xi_{a}-2 \over 8}{F^{2}_{a}}R^{2}+ {5 -4\xi_{a} \over 1024} n^{a} (R^{2})^{2}
+ {1 \over 256} n^{a} R^{4}) 
\nonumber\\
X^{a}_{10}(\xi_{a})&=&{i^{5} \over 5!}(F^{5}_{a}+ {5\xi_{a}-10\over 24} F^{3}_{a} R^{2} 
+ {10-5\xi_{a}^{2}\over 768}F_{a} (R^{2})^{2} 
+ {1 \over 96}F_{a} R^{4}) \quad .
\label{Xwz}
\ea
As mentioned above, the only ambiguities we find are parametrized by real 
constants $\xi_{a}$ that enter the definitions 
of the factors $X^{a}_{6}$, $X^{a}_{8}$ and $X^{a}_{10}$ in the anomaly polynomials
for $D=8,10$. Previous analyses \cite{asb} correspond to the choice $\xi_{a}=1$.
The parameters $\xi_{a}$ could be genuine free parameters in  
the effective lagrangian or an artifact of the procedure. 
In order to settle this issue one should compute additional string amplitudes 
on the disk. It is remarkable that these ambiguities are 
absent in the supersymmetric case and whenever $tr_{n^{a}} F = 0$ for 
any $a$, since all $\xi_{a}$ would disappear from the above formulas.  

The nice feature of (\ref{Xwz}) is that, one can easily recognize 
the closed string RR fields participating in the GSS mechanism of
anomay cancellation \cite{gs,as2}. Indeed the index $i$ labels the 
character\footnote{More precisely the index $i$ refers to the pair 
$(i, i^{C})$
of characters that are paired in the torus partition function.}
to which the corresponding closed string field belongs 
while $B^{i}_{a}$ measures the strength of the coupling 
of a ``generalized brane'' of type $a$ to these massless R-R states
and their massive (super-)partners.
Each sector includes $p+1$-form potentials of various degrees. The 
generalized GSS mechanism requires that a $p+1$-form 
potential $C_{p+1}^{i}$ in the sector labelled by $i$ couple 
with strength $B^{i}_{a}$ to a $D-p-1$-form $X^{a}_{D-p-1}$ of type $a$ in the 
factorized anomaly polynomial.
We can then extract from (\ref{Xwz}) all WZ couplings responsabile for this GSS 
mechanism in the effective $D$-dimensional theory  
\ba
L_{0} &=& 0
\\
L_{2} &=& {1\over 2} \sum_{{i,a}} B^{i}_{a} C_{0}^{i} X^{a}_{2}  
\\
L_{4} &=& {1\over 2} \sum_{{i,a}} B^{i}_{a} \left(
C_{0}^{i} X^{a}_{4} + C_{2}^{i} X^{a}_{2} \right)
\\
L_{6} &=& {1\over 2} \sum_{{i,a}} B^{i}_{a} \left( 
C_{0}^{i} X^{a}_{6}(1) + C_{2}^{i} X^{a}_{4} + C_{4}^{i} X^{a}_{2} \right) 
\\
L_{8} &=& {1\over 2} \sum_{{i,a}} B^{i}_{a} \left( 
C_{0}^{i} X^{a}_{8}(\xi) + C_{2}^{i} X^{a}_{6}(\xi) + C_{4}^{i} X^{a}_{4} 
+ C_{6}^{i} X^{a}_{2} \right) 
\\
L_{10} &=& {1\over 2} \sum_{{i,a}} B^{i}_{a} \left(
C_{0}^{i} X^{a}_{10}(\xi)+ C_{2}^{i} X^{a}_{8}(1) + 
C_{4}^{i} X^{a}_{6}(\xi)
+ C_{6}^{i} X^{a}_{4} + C_{8}^{i} X^{a}_{2} \right)
\label{wzc}
\ea
The factorization of the transverse channel 
amplitudes in the odd spin structure gives rise 
to the ``odd'' propagator which turns a $({p+1})$-form into its dual
$({D-p-3})$-form. More precisely 
\be
\langle d C_{p+1}^{i} d C_{D-p-3}^{j} \rangle = {\cal I}_{i} \Pi^{ij} 
{\cal V}_{D} \quad .
\label{propag}
\ee
with ${\cal V}_{D}$ a regularized D-dimensional volume.
Notice that one can simultaneously rescale any $C_{p+1}^{i}$ by a factor 
$\alpha_{p+1}^{i}$ and its dual $C_{D-p-1}^{i}$ by a factor 
$1/\alpha_{p+1}^{i}$ without modifying the factorization of the 
anomaly polynomial. The couplings (\ref{wzc}) will of course
absorb these renormalizations. 

From the expression of (\ref{wzc}) one can easily recognize the 
anomalous $U(1)$'s. They are to be identified as the abelian factors 
of type $a$ entering in $X^{a}_{2}$ and coupling anomalously to 
$C_{D-2}^{i}$ with a non-vanishing reflection coefficient
$B^{i}_{a}$. 

Although we have concentrated our attention on the CP-odd 
part of the effective lagrangian, in supersymmetric models some 
interesting CP-even coupling are related to (\ref{wzc}) by 
supersymmetry. WZ lagrangians in supersymmetric 
$D=4,6$ open string models have been worked it out in \cite{ss,rs}.
The terms displyed in (\ref{wzc}) cannot be the whole story.
RR fields belonging to sectors with vanishing Witten index are not 
present. In particular the WZ couplings RR axion belonging to the 
identity sector in any supersymmetric vacuum configurations in $D=4$ 
are not reproduced by (\ref{wzc}). In order to get around this problem
one has to find a way to disclose the anomalous content encoded in 
RR tadpoles
of sectors with vanishing Witten index.

\section{Tadpoles with ${\cal I} = 0$ and anomalous amplitudes}

In section 3 we have shown that RR tadpoles 
in sectors with non-vanishing Witten index are in one-to-one 
correspondence with irreducible anomalies. 
Unfortunately the vanishing of the  
Witten index ${\cal I}_i$ in a given internal sector makes the 
associated anomaly condition (\ref{quasi}) empty and the 
corresponding tadpole information lost. 
In this section we argue that a careful look into 
the anomaly structure of the effective theory may overcome
this apparent asymmetry.  
This will be done at the price of considering anomalous amplitudes involving
non-trivial insertions in the internal SCFT.   
The correct choice is given by the minimal number of insertions 
that removes the Bose-Fermi degeneracy in a given sector with ${\cal I} = 0$. 
As a result one can once again translate the associated tadpole condition
into a condition for anomaly cancellation through (\ref{quasi}).

Rather than being general let us illustrate this point in the simplest 
context of even-dimensional ($d=2\ell=10-D$) toroidal compactifications of the 
$SO(32)$ type I theory. To start with, we will neither 
turn on Wilson lines nor expectation values of the $B_{NS-NS}$-field 
(equivalent to non-commuting Wilson lines) \cite{bps}.
We will include their effects in the analysis later on.
In this
simple situation we have a single massless character ${\cal X}_0$.
Its Witten index
is clearly zero due to the $2\ell$ zero-modes 
$\Psi^{I}_0$ of the internal fermionic coordinates ($I=1...2\ell$). 
In order to get a non-trivial contribution from this sector
one should compute anomalous amplitudes involving 
insertions in the internal SCFT such that all fermionic zero-modes 
$\Psi^{I}_0$ are soaked up.
The contribution is now measured by non-vanishing generalized
Witten indices $\widehat{\cal I}_0 = Tr\, F^{\ell} (-)^F$
\footnote{The ``quasi'' topological index  
${\cal F} = Tr\, F (-)^F$ was 
introduced in the $N=2$ context in \cite{cfiv}}.
Irrespective of the dimension, anomalous amplitudes descending 
from the $D=10$ exagon meet the requirements. The string computation
is clearly the same as in ten dimensions but the
interpretation in terms of the lower dimensional field theory is 
different. With respect to the canonical anomaly,
the anomalous amplitudes under consideration are associated to higher derivative 
CP-odd terms involving scalars and KK vectors.
These receive anomalous contributions not only from 
loops of massless fermions and but also from their
higher KK-modes.     
The presence of such anomalous terms reflects the presence of a 
non-trivial coupling of the universal RR 2-form that compensates for
the anomalous variation of the fermion determinants only when $N=32$ as 
in $D=10$, \ie only when 
the RR tadpole of the identity sector with ${\cal I}_{0}=0$, is enforced.
If one turns on continuous Wilson lines, thus generically breaking 
$SO(32)$ to $U(1)^{16}$, one is effectively trading some 
anomalous contribution of the massless fermions to the CP-odd thresholds
with an additional contribution of the massive BPS states. 
If one turns on non-commuting Wilson lines, equivalent to a quantized 
background for the NS-NS antisymmetric tensor $B_{NS-NS}$, one is reducing both 
the rank of the Chan-Paton group as well as the number of gaugini and 
the coupling of the universal RR 2-form. 
Notice that this has a neat counterpart in the dual heterotic string.
In this case, modular invariance replaces the RR tadpole
condition in ensuring that any anomalous variation of 
the one-loop effective action is zero.
When a gauge symmetry $G$ in the heterotic string 
is realized in terms of a world-sheet current algebra at 
level $\kappa$ the effective GS 
coupling is rescaled by that same factor.
This is precisely what happens in the type I settings 
we have just discussed if 
one identifies the level of the worldsheet current algebra $\kappa$ 
with $2^{r/2}$ where $r$ is the rank of $B_{NS-NS}$.

In non-trivial open string configurations
involving sectors with vanishing Witten index ${\cal I}_i=0$,
other than the identity sector,
one can apply similar considerations.
The vanishing of ${\cal I}_i$ signals the presence of a
certain number $d$ 
of unsoaked fermionic zero-modes together with their bosonic 
superpartner under worldsheet supersymmetry. 
This essentially free piece of the internal SCFT is more like the 
$T^d$-toroidal compactification we have discussed so far.
Indeed one can trace the different higher dimensional origins of 
the relevant anomalous amplitudes by their 
different  scalings with the internal ``volumes''.
By volume here we do not neccesarily mean a geometrical one but
simply the result of the integration over the compact 
bosonic zero modes $X_0^I$ ($I=1,...d$).   
In the context of geometric orbifolds, different 
amplitudes\footnote{In our approach, amplitudes are linear combinations of 
characters. The former follow from the latter by inverting (\ref{chorb}).}
are in general associated to different fixed tori. The different 
scalings of the anomalous amplitudes allow one to identify the 
constraints on the different sets of branes and the corresponding  
RR tadpoles in sectors with ${\cal I}_i=0$.

It is instructive to explore the implications
of similar anomalous diagrams in different string settings.  
In vacuum configurations for the heterotic string,
moduli fields such as the radii of toroidal compactifications and the 
blowing up modes of an orbifold specify non-linear $\sigma$-models 
with continuous non-compact symmetries. These continuous symmetries 
combine geometric transformations such as internal diffeomorphisms
with non-geometric symmetries that arise because of 
the extended nature of the fundamental constituents. 
The related presence of winding states breaks
the continuous symmetries down to discrete ones (``T-duality'').
Some of these discrete symmetries of the tree-level effective 
lagrangian are broken by quantum effects.
On the one hand, they act by chiral transformations on the fermions 
that transform with moduli-dependent phases. 
The discrete charges (modular weights) are very sensitive to the 
sector each massless fermion belongs to. 
On the other hand, massive string states give rise to moduli-dependent 
thresholds corrections to CP-odd terms, 
related to the anomalous amplitudes discussed above,
in theories with fixed tori, \ie internal sectors with ${\cal I}_i = 0$. 
In ${\cal N}=1$ supersymmetric compactifications,
these CP-odd threshold corrections are anomalous in that they 
violate  an integrability condition \cite{dkl}. 
This violation is not renormalized beyond one-loop \cite{ant}. 
The combined variation of the contribution of the massless fermions 
and the CP-odd thresholds under discrete T-duality transformations is 
non-zero. There is howevere a universal GS counterterm that cancels the 
left-over discrete anomaly \cite{dkl}.
Worldsheet modular invariance at one-loop thus guarantees the absence
of target space modular anomalies.

After inclusion of all stable bound-states of strings and branes ``T-duality'' is 
expected to be promoted to ``U-duality''. 
In theories with open and unoriented strings, 
T-duality is broken at the perturbative level 
by the presence of the background brane configuration.
Still, by including the contribution of branes wrapping around cycles
of the internal manifold 
one expects to recover a discrete T-duality symmetry isomorphic to a 
T-duality subgroup of the U-duality of the parent theory. 
Absence of T-duality anomalies would then be a consequence of the 
vanishing of RR-tadpoles, including sectors with ${\cal I}=0$,
much in the same way as absence of 
T-duality anomalies is a consequence of modular invariance in 
closed-string theories. 

Following the same line of reasoning, one may argue that the absence 
of discrete anomalies in non-supersymmetric vacua requires the absence 
of all massless RR tadpoles. Although very little is known 
about the effective lagrangian in this 
case we expect the anomalous terms 
including the anomalous CP-odd thresholds to be ``topological'' and 
as such to be 
non-renormalized beyond one loop except possibly by world-sheet 
instantons or D-istantons. 

Let us conclude this section with a side remark on the odd-dimensional
case. Although there are no local anomalies for odd dimensional manifolds 
without boundaries and/or branes, global anomalies cannot be a priori 
excluded \cite{oddanomalies}. 
It is reassuring to observe that these are 
absent for toroidal compactifications with a Chan-Paton group of even 
rank as required by RR tadpole cancellation, \ie $N=32\times 2^{-r/2}$ 
or, for open descendants without open strings, $N=0$.

\section{Discussion and concluding remarks}

Most of the vacuum configurations with open and unoriented strings are 
not simply geometric compactifications of ten-dimensional theories
with open and unoriented superstrings. 
In a non-trivial background, genuine lower dimensional open
string excitations can be located at new kinds of 
branes that might not admit a large volume description.
Still one can try to extract some insights from the 
anomaly polynomial we have derived in some generality.
In principle, once the topological data of 
the ``compactification'' manifold are identified, one can 
extract some useful information about the relevant brane 
configuration and/or the vacuum gauge bundle by comparing 
the WZ couplings found in the genuine string computation with 
the ones that would result from a naive compactification. 
All additional terms are to be related to 
extra bound-states of branes and/or to non-trivial configurations of the 
background gauge fields.
We have parametrize the relevant information in terms of the boundary 
reflection coefficients $B^{i}_{a}$, \ie one-point functions of 
massless states of type $(i,i^{C})$ on the disk with boundary 
conditions of type $a$. These $B^{i}_{a}$ enjoy some interesting 
completeness properties and behave as vielbeins, \ie they allow one to 
transform an index of type $i$ labelling a sector of the closed string 
spectrum that flows in the transverse channel into an index of type 
$a$ that labels the independent Chan-Paton multiplicities.
It is conceivable that the $B^{i}_{a}$ satisfy some interesting 
evolution equation that would allow one to compute them starting 
from special ``rational'' points in the moduli space of the vacuum 
configuration.

At these points, classifying all possible boundary conditions compatible 
with preserving half of the bulk symmetries is expected to be 
feasible.  The worldsheet SCFT description 
pursued in this paper and 
the topological nature of the anomaly-related terms should be 
sufficient to show that the $B^{i}_{a}$ for massless sectors with 
non-vanishing Witten index should be independent of the moduli.
Indeed, for D-branes wrapped around supersymmetric cycles in Calabi-Yau
spaces it is known that $B^{i}_{a}$ for $i$ a massless sector 
are quasi-topological \cite{ooy}. 
For A-type boundary conditions the $B^{i}_{a}$, with $i$ running over the 
(c,c) primary fields and $a$ over the middle cohomology cycles, only depends 
on the complex structure deformations and may be computed at large 
volumes. For B-type boundary conditions the $B^{i}_{a}$, with $i$ running over 
the 
(c,a) primary fields and $a$ over the vertical cohomology cycles, only depends 
on the K\"ahler structure deformations and can be computed in terms of 
the ``quantum intersection form'', \ie the ``topological intersection 
form'' (an integer) plus worldsheet instanton corrections. 
For massive sectors this is no longer expected to be the case, but 
the polynomial equations satisfied by the $B^{i}_{a}$ may rescue the 
situation \cite{fpss}.
It would be interesting to find a way to extend the arguments of  \cite{ooy}
based on the topological twist of ${\cal N}=2$ SCFT to worldsheet 
theories that only enjoy ${\cal N}=1$ superconformal symmetry. 

We have not discussed the tadpoles in the NS-NS sector that may be 
disposed of by means of a Fischler-Susskind mechanism. The latter
may destabilize the vacuum configuration. A signal could be found in 
behaviour of the $B^{i}_{a}$ at points where a given brane 
configuration could decay into its constituents or tachyon 
condensation could trigger the formation of a stable non-BPS bound 
state.  

The applications of the anomaly-tadpole correspondence we have 
established go far beyond of the contexts exploited in this paper. 
Indeed, our results only rely on the consistency between the open and
closed string channel in the presence of
boundaries and crosscaps. Relying on the ${\cal N}=(1,1)$
superconformal symmetry on the worldsheet enables one to isolate the odd 
spin-structure as the source of all potential anomalies. Modular 
invariance then allows one to identify anomalous amplitudes with RR tadpoles. 
Similar analyses can be applied to any
(BPS or not) brane configuration satisfying the consistency requirements.
Tadpoles are associated to closed string absortion processes
encoded in the dynamics of boundaries and crosscaps on the worldsheet,
Anomalies enter as one-loop CP-odd effects in the effective theory 
governing the dynamics of the brane configuration.
The UV-IR dictionary relating these two effects is expected to be realized 
in any SYM/SUGRA correspondence described in terms of boundaries
on the string worldsheet.

\vspace*{1cm}

{\large{\bf Acknowledgements}}

\vspace*{0.3cm}
\noindent
We would like to thank A.~Sagnotti and Ya.S.Stanev for valuable 
discussions, M. Serone for collaboration at early stages of this project
and G. Pradisi and G.C. Rossi for useful comments.

\end{document}